\documentclass[astra]{copernicus}
\usepackage{hyperref}

\begin{document}

\title{Semi-analytical model of cosmic ray electron transport}

\author{A. Ivascenko}
\author{F. Spanier}

\affil{Universit\"at W\"urzburg, ITPA, Lehrstuhl für Astronomie, Emil-Fischer-Str. 31, 97074 W\"urzburg, Germany}

\runningtitle{CR electron transport model}

\runningauthor{A. Ivascenko and F. Spanier}

\correspondence{A. Ivascenko\\ (aivascenko@astro.uni-wuerzburg.de)}

\received{12 November 2010}
\revised{25 January 2011}
\accepted{14 February 2011}
\published{}

\firstpage{1}

\maketitle

\begin{abstract}
    We present a numerical extension to the analytical propagation model introduced in \citet{hein08} to describe the leptonic population in the galactic disc. The model is used to derive a possible identification of the components that contribute to the leptonic cosmic ray spectrum, as measured by PAMELA, Fermi and HESS, with an emphasis on secondary $e^+-e^-$ production in collisions of cosmic ray particles with ambient interstellar medium (ISM). We find that besides secondaries, an additional source symmetric in $e^+$ and $e^-$ production is needed to explain both the PAMELA anomaly and the Fermi bump, assuming a power-law primary electron spectrum. Our model also allows us to derive constraints for some properties of the ISM.
\end{abstract}

\introduction
Recently the leptonic component of the cosmic ray spectrum has gained new attention. New observations from ATIC, PAMELA and Fermi show a deviation from a power-law in the form of an excess in both the electron and positron spectra. Annihilating dark matter \citep{allahverdi08} and nearby pulsars \citep{busching08,grimani2009,amatoblasi2010} (among other hypotheses) have been proposed as possible sources of the excess leptons. Regardless of the source, a propagation model is needed to connect the energy spectrum measured on Earth with the injection \mbox{spectra.}

    We present our numerical cosmic ray transport model in application to the high energy electron transport in the ISM. Spatial and momentum diffusion, particle escape, acceleration via Fermi I and continuous energy losses were taken into account and their effects on the steady-state energy spectrum analyzed. In solving the transport equation we employed quasi-linear transport theory, the diffusion approximation and a separation of the spatial and momentum problem to obtain the leaky-box-equation, which was then solved numerically. The spatial problem was solved analytically in cylindrical and prolate spheroidal coordinates.

    The transport model was employed to calculate the spectrum of secondary electrons in our galaxy. We assume the leptons from pioArXivn decay to be the dominating component of the secondary spectrum. The lepton spectrum from pion collisions of highly relativistic cosmic ray protons with thermal protons in the ISM was calculated using the parametrization of pion production in p-p-collisions from \citet{Aharonian} and used as the injection spectrum for the transport model. With realistic simulation parameters the resulting positron flux in the vicinity of the solar system lies remarkably close to the low-energy PAMELA data. Assuming a generic power-law primary $e^-$ spectrum and an additional $e^+-e^-$ source allows us to fit the datasets from PAMELA, Fermi and HESS very nicely and to put constraints on several transport model parameters and the corresponding ISM properties.

\section{The data}
The leptonic cosmic ray spectrum gained new attention when the results of the balloon experiment ATIC were published \citep{aticfeature}. The authors claimed an anomalous excess in the spectrum just below 1\,TeV. This started wild speculations about the nature of the source for these leptons, which should be in the relative vicinity of the solar system.

Those speculations were fueled further by new measurements from the PAMELA satellite published in \citet{pamelapositron} claiming a rise in the positron spectrum above 10 GeV. The lepton spectrum from the Fermi satellite \citep{fermielectron} could not confirm the ATIC-peak quantitatively but showed a deviation from a straight power-law in the same energy region.

The HESS atmospheric Cerenkov telescope extended the spectrum measurement beyond 1\,TeV showing a cut-off or a steeper power-law but could also not confirm the ATIC excess \citep{hesselectron}.

Recently the Fermi lepton spectrum was extended to lower energies (7\,GeV) in \citet{fermilowen}. Also the PAMELA results were updated with new experimental data and different evaluation techniques in \citet{pamelaneu}.

To interpret these measurements, it is necessary to connect the injection spectra at the (possible) sources with the observations in the solar system. To do that we need a transport model that includes all the relevant processes in the ISM (spatial and momentum diffusion, cooling, particle escape, etc.).

\section{Transport model}
The derivation of the transport equation follows closely \citet{LS88} and \citet{hein08}. Using Quasi Linear Theory and the diffusion approximation the Vlasov equation is transformed into the steady-state diffusion-convection equation for the phase space density $f(\mathbf{x}, p)$, which can be written in the form
    \begin{equation}
        \mathcal{L}_x f + \mathcal{L}_p f + S(\textbf{x},p) = 0
    \end{equation}
    with the spatial operator
    \begin{equation}
        \mathcal{L}_x(\textbf{x},p) \equiv \nabla \left[\kappa(\textbf{x},p) \nabla \right]
    \end{equation}
    and the momentum operator
    \begin{equation}
        \mathcal{L}_p(\textbf{x},p) \equiv \frac{1}{p^2} \frac{\partial}{\partial p} \left( p^2 D(\textbf{x},p) \frac{\partial}{\partial p} - p^2 \dot{p}(\textbf{x},p) \right) - \frac{1}{\tau(\textbf{x},p)} \textrm{.}
    \end{equation}
    This equation describes the diffusion (coefficients $\kappa$ and $D$ in space and momentum, respectively), acceleration and cooling (rates $\dot{p}$) and escape (timescale $\tau$) of an injected source distribution $S$.

    Following \citet{LS88} the equation can be solved by separating the operators and the source term in their spatial and momentum dependencies. Then the resulting steady state particle distribution can be written as an infinite sum:
    \begin{equation}
        f(\textbf{x},r) = \sum_i A_i(\textbf{x}) R_i(p)
    \end{equation}
    The spatial coefficients $A_i(\textbf{x})$ can be acquired by solving the diffusion equation in the appropriate geometry
    \begin{equation}
        \frac{\partial}{\partial u} A_i \exp(-\omega_i ^2 u)  = \nabla \left( \kappa \nabla A_i \exp(-\omega_i ^2 u) \right) \textrm{,}
    \end{equation}
    where $\omega_i ^2$ take the role of inverse escape times for the momentum modes and $u$ is a convolution variable.

    In this work we used the analytical solution for cylindrical geometries as provided by \citet{wangschlick87} since this offers the best reflection of the symmetries of a spiral galaxy. Analytical solutions in spherical and prolate spheroidal coordinates are also available \citep{SchlickSievers87, hein08} and can be used to describe particle transport in elliptical galaxies and galaxy clusters.

    The momentum modes $R_i(p)$ are solutions of the ODE, commonly referred to as the leaky-box equation:
    \begin{equation}
        \frac{1}{p^2} \frac{\partial}{\partial p} \left[ \frac{a_2 p^4}{\kappa(p)} \frac{\partial R_i}{\partial p} - \frac{a_1 p^3}{\kappa(p)} R_i + \dot{p} R_i \right] - \omega_i^2 \kappa(p) R_i = -Q(p)
    \end{equation}
    The parameters $a_1$ and $a_2$ represent the absolute scales for Fermi I and II processes and depend mainly on the Alfven and shock speed, respectively.
    As \citet{LS88} have shown, an analytical solution for the equation can be found, if the cooling rate $\dot{p}$ scales linearly with the particle energy, which is a good assumption for high energy protons and heavier nuclei. Electrons above a few GeV, however, are cooled primarily by synchrotron radiation and IC scattering, both scaling quadratically in particle energy. Therefore the leaky-box equation was solved using numeric relaxation methods.

    To summarize, the model describes a particle distribution that is injected by a homogeneous population of time-independent sources in the galactic disc (injection region), diffuses throughout the galactic disc and halo (confinement region), cooling by bremsstrahlung, adiabatic deceleration, synchrotron radiation and inverse Compton scattering, and leaves the confinement region at some point. The matter, photon and magnetic field distributions are assumed to be homogeneous over the whole confinement region.

    A model with a uniform source distribution is not a good choice to simulate high energy leptons from discrete sources (SNR, PWN\ldots), since the short energy loss timescales make the source distribution critical for the resulting spectra, as discussed in \citet{amatoblasi2010}. Conversely, production of secondaries from p-p~collisions in the galactic halo can be approximated very well as a homogeneous source, since there are no dramatic variations in the CR and ISM densities.

    A similar analysis by \citet{moskalenkostrong98} used an analogue model as in this paper, but obviously did not include the new data above 1\,GeV. \citet{stephens2001} has basically the same drawbacks, in addition this work has a more sophisticated secondary production model. Compared with the recent works in this area like \citet{amatoblasi2010}, we concentrate more on the low energy end of the current PAMELA data and the constraints it can put on propagation parameters if interpreted as secondaries.

\section{Secondary leptons}
Fortunately there is a lepton production process that can be approximated with a homogeneous spatial distribution: the secondary leptons from the decay of products of proton collisions. To get a galaxy-averaged flux of secondaries we consider collisions of high energy cosmic ray protons with ambient thermal matter, assuming that both populations are homogeneously distributed with average densities.

    The pion production cross-section and the electron source function for a single p-p collision used here are parametrizations of results from the SIBYLL event generator from \citet{Aharonian}. Folding the source function $F_e \left( \frac{E_e}{E_p},E_e \right)$ with the cosmic ray spectrum $J_p(E_p)$ and the inelastic collision cross-section $\sigma_{\mathrm{inel}}(E_p)$ yields the electron flux:
    \begin{equation}
        \frac{dN_e}{dE} = 4\pi n_H \int \sigma_{\mathrm{inel}}(E_p) J_p(E_p) F_e \left( \frac{E_e}{E_p},E_e \right) \frac{dE_p}{E_p}
        \label{sec_flux}
    \end{equation}
    Since the source function is sharply peaked, the CR spectrum can be cut off exponentially at the first knee ($\approx 1000$ TeV) with negligible deviations to the lepton spectrum below 10 TeV:
    \[ J_p(E_p) = 0.252 \left( \frac{E_p}{\mathrm{TeV}} \right)^{-2.677} e^{\frac{-E_p}{10^3\ \mathrm{TeV}}} \mathrm{m}^{-2} \mathrm{sr}^{-1} \mathrm{s}^{-1} \mathrm{TeV}^{-1} \]
    This parametrisation of the CR spectrum was obtained by fitting the CR data between $10^{10}$ and $10^{14}$ eV in \citet{cronin97}.

    The average thermal matter density $n_H$ is treated as a free parameter. The resulting electron flux with a slightly harder power-law (spectral index $2.62$) is used as the injection function in the transport model. Since the parametrizations in \citet{Aharonian} only consider p-p~interactions, the actual flux of secondaries is underestimated by the contribution from heavier nuclei.

\section{Parameter constraints}

\begin{figure}[t]
\vspace*{2mm}
\centering
\includegraphics[width=8.3cm]{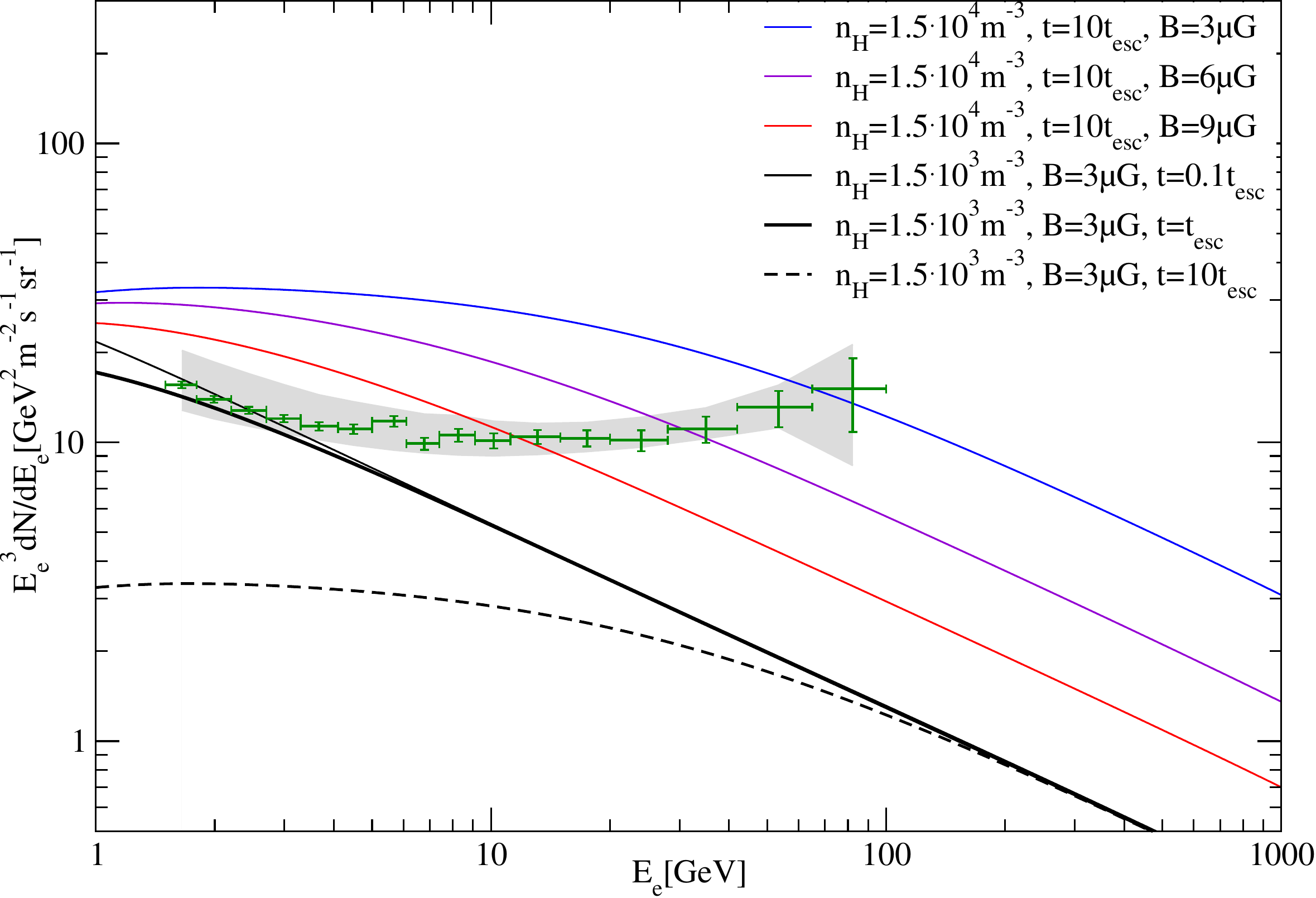}
\caption{Parameter studies for the secondary positron spectrum compared to the PAMELA $e^+$ fraction multiplied by a fit of the Fermi $e^+ +e^-$ data. Combined systematic errors of PAMELA and Fermi are shown by the grey area. The thick black line with $n_H=1.5\cdot10^3\,\mathrm{m}^{-3}$, $B=3\,\mathrm{\mu G}$ and $t=t_{\mathrm{esc}}$ yields the best fit of the positron spectrum, if combined with an extra high energy source as shown in Fig.~\ref{fig:fit}. The thin and dashed black lines show the slope change with changing escape losses.}
\label{fig:parameter}
\end{figure}

The principal shape of the resulting steady-state lepton spectrum is shown in Fig.~\ref{fig:parameter}. The cooling, diffusion and escape processes introduce breaks in the injected power-law, so that multiple energy intervals with different spectral indices are formed, each of them defined by a different process. At the first break the injected power-law is steepened by $1/3$ as the diffusion/escape time scale becomes shorter than the time scale of bremsstrahlung and adiabatic cooling (linear in energy). The second break appears when the time scale for synchrotron and IC losses becomes shorter that the diffusion time scale. These quadratic losses steepen the spectrum by 1. The absolute flux value is determined mainly by the ISM density, since the injected secondary spectrum has a linear $n_H$ dependence (Eq.~(\ref{sec_flux})), and to a lesser extent by escape and cooling losses, which transport particles out of the simulation region.

So the model has three major parameters: the linear energy loss time scale that reflects the thermal gas density $n_H$, the quadratic loss time scale that incorporates the energy densities of the EM field (composed of a constant CMB density of $\approx 0.3$\,eV/cm$^3$ and a variable magnetic field $B$) and the escape time scale $t_{\mathrm{esc}}$ that corresponds to the size of the confinement region.

Measurements of the positron spectrum, in particular the newest results from PAMELA, allow us to constraint those parameters. A rather conservative statement can be made, if we assume that the secondaries have a negligible contribution to the positron spectrum. As shown in Fig.~\ref{fig:parameter}, assuming an average ISM density of $\approx10^3$\,m$^{-3}$, the confinement region has to be less than the assumed $30$\,kpc or the diffusion process has to be more efficient, so that the particles leave the galaxy 10 times faster (black dotted line). A higher ISM density would require an even more efficient particle escape, so as not to overshoot the measured positron spectrum. Since the spectral shape wouldn't play any role, the magnetic field can be almost arbitrary.

\begin{figure*}[t]
\vspace*{2mm}
\centering
\figbox*{}{}{{\includegraphics[width=12cm]{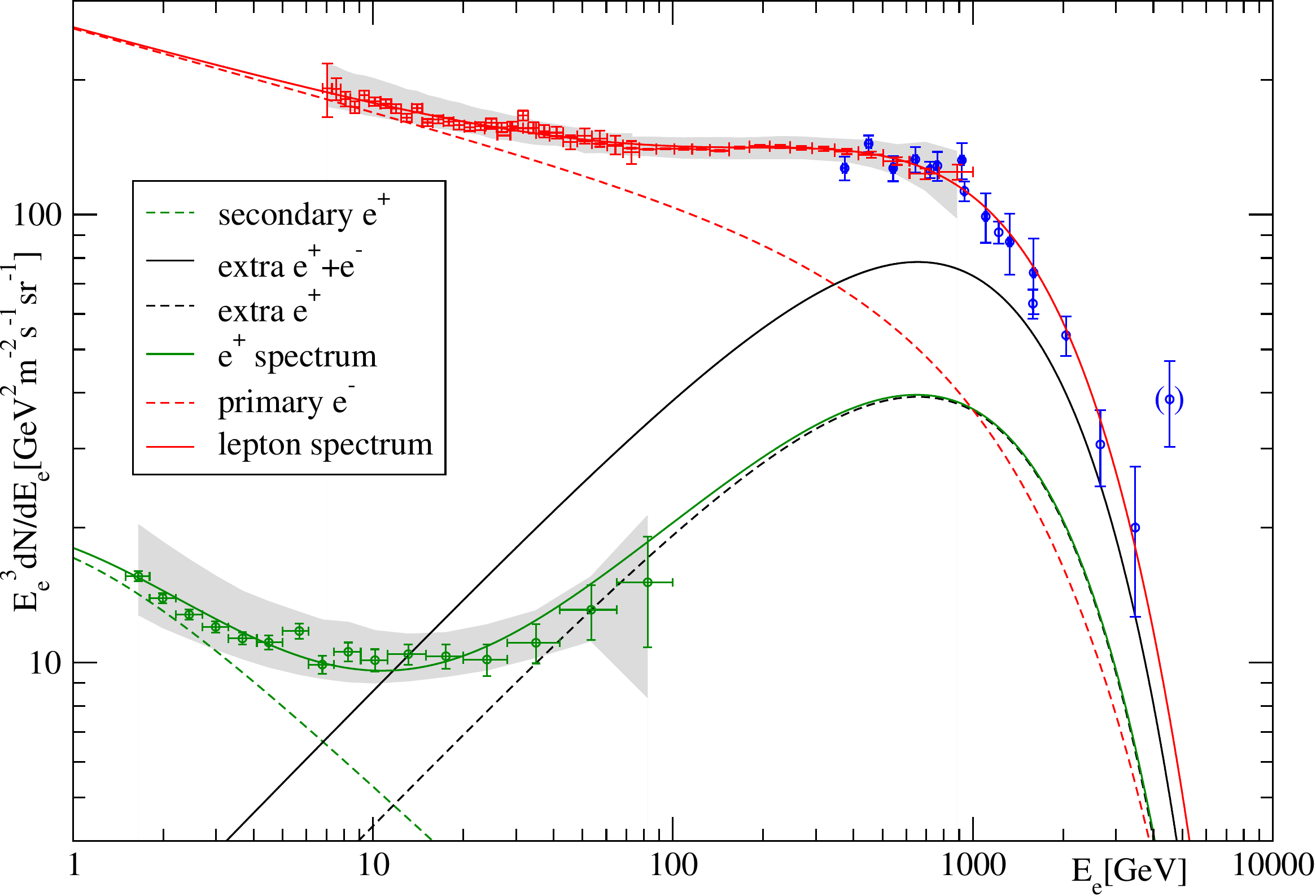}}}
\caption{A fit of Fermi \citep{fermilowen} and HESS \citep{hesselectron} lepton data (solid red line) and PAMELA \citep{pamelaneu} positron data (solid green line). Grey areas show the systematic errors. The components that contribute to the spectra are secondary positrons (dashed green), a generic power-law primary spectrum (dashed red) and an additional symmetric $e^+ -e^-$ source (solid and dashed black).}
\label{fig:fit}
\end{figure*}

On the other hand, the PAMELA data can be fitted very well with a superposition of two power-laws with a spectral index of about $3.6$ for the low energy part, corresponding remarkably well with the secondary spectrum steepened by synchrotron and IC losses. Assuming the low energy positrons to be secondaries puts much stricter constraints on the parameters, since the second break in the spectrum has to appear at $\approx 1$ GeV to match the data. The break energy is determined by the equilibrium of the diffusive/escape losses and the radiative losses. The thick black line in Fig.~\ref{fig:parameter} is the best fit with $n_H=1.5\cdot 10^3$\,m$^{-3}$, $B=3\,\mu G$ and $t=t_{\mathrm{esc}}$. The radiative losses in the galactic halo are dominated by IC scattering on the CMB photons, so that lowering the magnetic field has no effect on the break energy. A slight flattening at the low energy end of the spectrum is needed for the best fit of the new PAMELA data. This corresponds to a halo radius of $30$ kpc and a spatial diffusion coefficient $\kappa(E=1\,\mathrm{GeV}) = 4.5\cdot10^{24}$\,m$^2$\,s$^{-1}$, yielding an average confinement time $t_{\mathrm{esc}}=2\cdot10^{17}$\,s. The absolute flux fixes the third parameter, the ISM density, to $1.5\cdot10^3$\,m$^{-3}$. Taking into consideration that about $99\%$ of the confinement region represent the galactic halo with an average gas density of about $10^3$\,m$^{-3}$ \citep{mckee}, the parameters seem very realistic.

\section{Lepton excess}
In Fig.~\ref{fig:fit} the latest Fermi lepton measurement, in which the spectrum has been extended down to $7$\,GeV, is shown. A generic fit of it was used to calculate a positron flux from the PAMELA positron fraction measurement. This data is used because it was taken simultaneously in the same solar cycle making interpretation easier. The HESS high-energy data was shifted down by $15\%$ to better coincide with the Fermi data. This is still within the systematic error margins of HESS, not to mention the systematic error of Fermi being of the same order of magnitude.

Once the secondary background has been subtracted from the PAMELA positron flux, the excess can be fitted very well with a simple power-law with a spectral index of $2.3$ (Fig.~\ref{fig:fit}, black dashed line). Remarkably, if we assume the source of these ``extra'' positrons to be symmetric in $e^+ -e^-$, we can also fit the Fermi bump leaving a primary lepton component of the shape $E^{-3.2}\cdot\exp(-E/1.46$\,TeV) (red dashed line). The (shifted) high-energy HESS data provides a means to fix the cut-off energy of the ``extra'' leptons to about 920\,GeV.


\conclusions
With the help of our semi-analytic transport model we were able to identify the low-energy PAMELA data as secondary positrons from collisions of CR protons with ISM gas using very realistic values for the simulation parameters. An indirect implication from that is that lepton transport is dominated by radiative losses (synchrotron and IC) above 1\,GeV.

Assuming a generic primary lepton spectrum of the form $E^{-3.2}\cdot\mathrm{exp}(-E/1.46$\,TeV), the PAMELA anomaly, the Fermi bump and the high energy HESS data can all be fitted by a single component, that is symmetric in $e^+ -e^-$ with a power-law spectral index of $2.3$ and a cut-off energy of 920\,GeV. This could either be a lepton population injected into the ISM by nearby source with $E^{-2.3}$, so that the propagation time scale is shorter than the radiative loss time scale, or a more distant source with a much harder spectrum $E^{-1.3}$ which is then steepened by radiative losses. A supernova remnant would be a good candidate in the first case, pulsars of different ages would suite both cases.

\begin{acknowledgements}
We acknowledge the support by the Deutsche Forschungsgesellschaft through the Graduate School ``Astroplasmaphysik'' and through grant SP 1124-3.\\~\\
Edited by: R.\ Vainio\\
Reviewed by: C. Grimani and another anonymous referee
\end{acknowledgements}

\end{document}